\documentclass[11pt]{article}
\usepackage{moriond,epsfig}

\def\Journal#1#2#3#4{{#1} {\bf #2}, #3 (#4)}

\def\NIMA{{\em Nucl. Instrum. Methods} A}

\def\PRL{\em Phys. Rev. Lett.}
\def\PRD{{\em Phys. Rev.} D}

\def\be{\begin{equation}}
\def\ee{\end{equation}}
\def\bea{\begin{eqnarray}}
\def\eea{\end{eqnarray}}

\newcommand{\numutonutau}{$\nu_{\mu}\rightarrow\nu_{\tau}$ }
\newcommand{\numutonue}{$\nu_{\mu}\rightarrow\nu_{e}$ }

\newcommand{\numu}{$\nu_{\mu}$}
\newcommand{\numubar}{$\overline{\nu}_{\mu}$}
\newcommand{\nue}{$\nu_{e}$}
\newcommand{\nuebar}{$\overline{\nu}_{e}$}

\newcommand{\cs}{$\chi^{2}$}

\newcommand{\evsq}{${\rm eV}^{2}/c^{4}$}

\newcommand{\nutau}{$\nu_{\tau}$}
\newcommand{\dmsq}{$|\Delta m^{2}_{32}|$}

\newcommand{\sintwo}{$\rm sin^{2}(2\theta_{23})$}

\newcommand{\ket}[1]{|#1\rangle}

\begin{document}
\vspace*{4cm} 
\title{ \mbox{MINOS} RESULTS, PROGRESS AND FUTURE PROSPECTS} 
\author{\sc TOBIAS M. RAUFER, for the \mbox{MINOS} Collaboration}
\address{University of Oxford, Sub-dept. of Particle Physics\\ 
  Denys Wilkinson Building, Keble Road \\ 
  Oxford OX1 3RH. United Kingdom}
\footnotetext{Correspondence: \tt t.raufer1@physics.ox.ac.uk}

\maketitle\abstracts{ The MINOS long baseline experiment has been
collecting neutrino beam data since March 2005 and has accumulated
$3\times10^{20}$ protons-on-target (POT) to date. MINOS uses Fermilab's
NuMI neutrino beam which is measured by two steel-scintillator tracking
calorimeters, one at Fermilab and the other 735\,km downstream, in
northern Minnesota. By observing the oscillatory structure in the
neutrino energy spectrum, MINOS can precisely measure the neutrino
oscillation parameters in the atmospheric sector. From analysis of the
first year of data, corresponding to $1.27\times10^{20}$\,POT, these
parameters were determined to be \dmsq~=~2.74$^{+0.44}_{-0.26}\times
10^{-3}$\,\evsq{} and \sintwo~$>$ 0.87 (68\% C.L.). MINOS is able to
measure the neutrino velocity by comparing the arrival times of the
neutrino beam in its two detectors. Using a total of 473 Far Detector
events, $(v-c)/c = (5.1 \pm 2.9)\times10^{-5}$ (68\% C.L.) was
measured. In addition, we report recent progress in the analysis of
neutral current events and give an outline of experimental goals for the
future.  }

\section{Introduction}
It is now well established that neutrinos have non-zero masses and that
neutrinos mix. Their weak interaction eigenstates (or ``flavour''
eigenstates) $\nu_{\alpha}$ are related to their mass eigenstates
$\nu_i$ by a unitary transformation $U$:
\begin{equation}
\ket{\nu_{\alpha}} = \sum_i U^{*}_{\alpha i} \ket{\nu_{i}}
\end{equation}
$U$ is called the PMNS \cite{pontecorvo,mns} matrix. Neutrinos are
created and detected by weak interaction processes but their propagation
in free space is described by their mass eigenstates, causing relative
phases to change. This leads to the
phenomenon of neutrino oscillations.

MINOS is a long baseline neutrino oscillation search based at FNAL. The
neutrino beam created in the NuMI beamline is sampled in two locations,
$\sim$1\,km from the beam production target underground at Fermilab, and
again $735$\,km downstream in the Soudan Underground Laboratory, in
Minnesota. 

Using this setup, MINOS measures the oscillation parameters \dmsq and
 \sintwo to world leading precision. In addition, MINOS searches for
 sub-dominant \numutonue oscillations, oscillations into sterile
 neutrinos, and $\nu\rightarrow\overline{\nu}$ transitions.  The MINOS
 Far Detector can also be used to detect neutrinos created in the
 atmosphere. \cite{Adamson:2005,Adamson:2007}

\section{Experimental setup}
\subsection{The NuMI beamline} 
\begin{figure}
  \begin{center}
    \includegraphics[width=\textwidth]{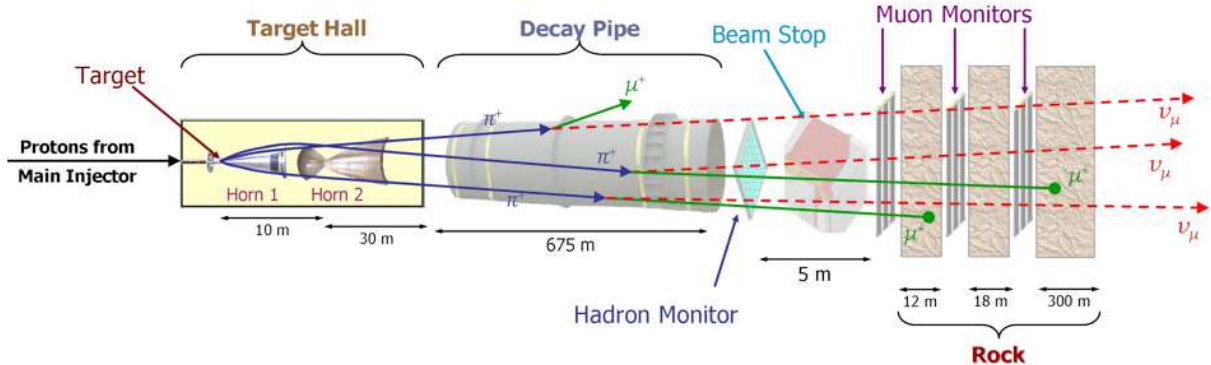}
    \caption{The NuMI beamline components. Positively charged mesons produced
    in the target are focused by two magnetic horns and subsequently
    decay in a 675\,m-long evacuated volume producing predominantly
    $\nu_\mu$. The remaining hadrons are stopped in a beam absorber at
    the end of the decay pipe. 
      \label{fig:NuMI}}
  \end{center}
\end{figure}

A schematic of the NuMI neutrino beamline is shown in Figure
\ref{fig:NuMI}. NuMI uses protons with a momentum of 120\,GeV extracted
from Fermilab's Main Injector accelerator. They impinge a 95.4\,cm long
segmented graphite target producing secondary particles, mainly $\pi$
and $K$ mesons. The positively charged mesons are focused using a system
of two pulsed, parabolic magnetic focusing elements, called
``horns''. The secondaries subsequently decay in a 675\,m long, 2\,m
diameter evacuated decay volume producing neutrinos. 

Hadrons reaching the end of the decay volume without decaying are
stopped in an beam absorber following the decay pipe. The beam absorber
consists of a water-cooled aluminium core surrounded by a layer of steel
blocks and an outer layer of concrete. The remaining muons from the
meson decays are stopped in the $\sim$300\,m of rock that separate the
MINOS Near Detector hall from the beam absorber.

The target position relative to the first horn and the horn current
are variable. For most of the collection of data used in the
oscillations analysis presented here, the target was inserted 50.4\,cm
into the first horn to maximize neutrino production in the 1-3\,GeV
energy range.  The data described here were recorded in this position,
between May 2005 and February 2006, and correspond to a total of
1.27$\times 10^{20}$ POT.  The charged current (CC) neutrino event
yields at the ND are predicted to be 92.9\% \numu, 5.8\% \numubar, 1.2\%
\nue{} and 0.1\% \nuebar.

\subsection{The MINOS detectors}
The MINOS detectors are designed to be as similar as possible while
operating in very different conditions. They are steel-scintillator
sampling calorimeters, magnetized to $\sim$1.3\,T allowing to measure
particle momentum using track curvature as well as range. The
scintillator planes are divided into 4.1\,cm wide strips rotated by
$90^{\circ}$ on subsequent planes to enable 3-dimensional event
reconstruction. The scintillator light is captured by embedded
wavelength shifting (WLS) fibres and transported to the edge of the
detectors, where the optical signal is converted using Hamamatsu M64
\cite{PMTs} (Near Detector) and M16 \cite{PMTs2} (Far Detector)
photomultiplier tubes.

Below 10\,GeV, the hadronic energy resolution was measured to be
56\%/$\sqrt{E\rm{[GeV]}}\oplus 2\%$ and the EM resolution was measured
to be 21.4\%/$\sqrt{E\rm{[GeV]}}\oplus 4.1\%/E\rm{[GeV]}$.\ \cite{caldet}
The muon energy resolution $\Delta E_{\mu}/E_{\mu}$ varies smoothly from
6\% for E$_\mu$ above 1 GeV where most tracks are contained and measured
by range, to 13\% at high energies, where the curvature measurement is
primarily used.

\section{Monte Carlo tuning}
\label{sec:beamTuning}
\begin{figure}[!htb]
\begin{center}
  \includegraphics[width=0.7\textwidth]{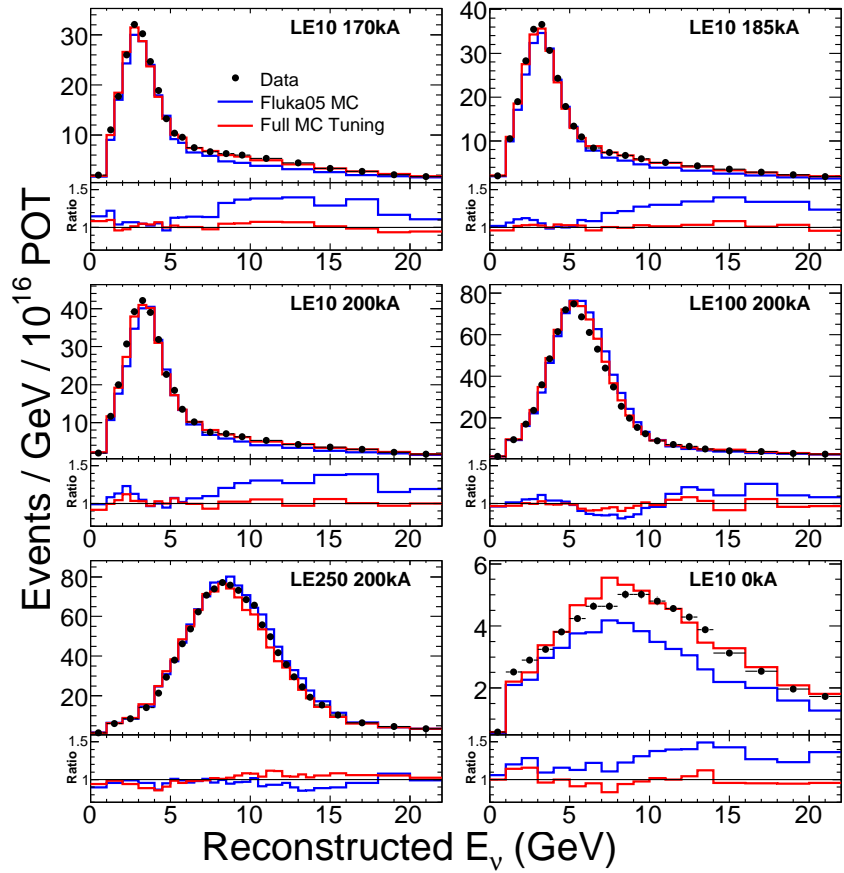}
  \caption{ Data and Monte Carlo energy spectra and their ratios are
  shown for six deferent beam configurations. The blue line corresponds
  to the untuned, the red line to the tuned hadron production model. The
  tuning significantly improves data/MC agreement in all cases.  }
  \label{fig:beamTuning}
  \end{center}
\end{figure}

MINOS took data with 6 beam configurations obtained by varying the
target position and the current in the focusing horns. The Monte Carlo
simulations of neutrino fluxes strongly depend on the underlying models
of hadron production in the target, which are presently poorly
constrained at MINOS energies. The nominal simulations, based on {\tt
FLUKA05}, yielded energy spectra which did not match the high-statistics
data in the Near Detector. A better agreement was achieved by smoothly
adjusting the $p_z$ and $p_T$ of hadrons (mostly pions) produced off the
graphite target. The resulting spectra, which describe the data much
more closely than the nominal simulations, are shown in Figure
\ref{fig:beamTuning}.

\section{\numu\ disappearance analysis}
\subsection{Event selection and extrapolation}
In the Far Detector (FD), the signal from eight scintillator strips is
read out by the same PMT pixel. Therefore, the initial step in the
reconstruction of the FD data is the removal of the eightfold
hit-to-strip ambiguity using information from both strip ends. In the
ND, timing and spatial information is first used to separate individual
neutrino interactions from the same spill. Subsequent reconstruction is
done in the same way in both detectors. Tracks are found and fitted, and
showers are reconstructed to be combined to events. For \numu{} CC
events, the total reconstructed event energy is obtained by summing the
muon energy and the visible energy of the hadronic system.  
To prevent human biases when assessing the oscillation analysis results,
a blinding mechanism was applied to the FD data set.  This procedure hid
a substantial fraction of the FD events with the precise fraction and
energy spectrum of the hidden sample unknown. Events are pre-selected in
both detectors, by requiring total reconstructed energy below 30\,GeV
and a negatively charged track.  The track vertex must be within a
fiducial volume such that cosmic rays are rejected and the hadronic
energy of the event is contained within the volume of the detector.  The
pre-selected \numu{} event sample is predominantly CC with a 8.6\%
neutral current (NC) event background estimated from Monte Carlo (MC)
simulations.  The fiducial mass of the FD and ND is 72.9\% and 4.5\% of
the total detector mass respectively.

\begin{figure} [!htb]
\begin{center}
\includegraphics[width=0.55\textwidth]{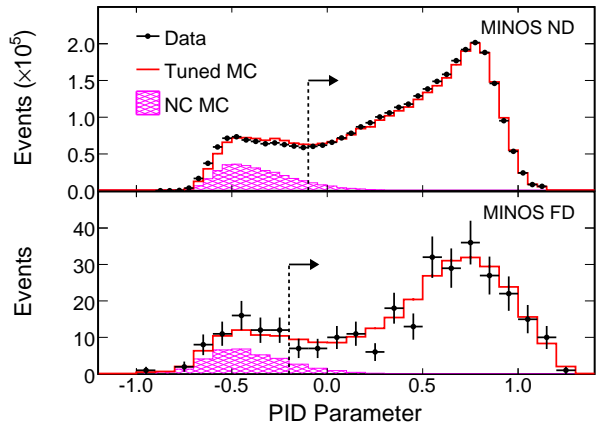}
\caption{\label{fig:pidn} Data and tuned MC predictions for the PID
  variable in the ND (top) and FD (bottom). The arrows depict the
  positions of the selection cuts.  The FD MC distribution for CC events
  uses the best fit parameters discussed in the text.}
\end{center}
\end{figure}

A particle identification parameter (PID) incorporating probability
density functions for the event length, the fraction of energy contained
in the track and the average track pulse height per plane provides
separation of \numu{} CC and NC events.  The PID is shown in
Figure\,\ref{fig:pidn} for ND and FD data overlaid with simulations of
NC and CC events.  Events with PID above -0.2 (FD) and -0.1 (ND) are
selected as being predominantly CC in origin. These values were
optimized for both detectors such that the resulting purity of each
sample is about 98\%.  The efficiencies for selecting \numu{} CC events
in the fiducial volume with energy below 30\,GeV are 74\% (FD) and 67\%
(ND).

The measurement of the energy spectrum at the ND is used to predict the
unoscillated spectrum at the FD.  Fits to the ND data yield tuning
parameters for the predicted neutrino flux. These fits are based on
parameterisations of the secondary pion production at the NuMI target as
a function of $x_{F}$ and $p_{T}$ as described in Section
\ref{sec:beamTuning}. The FD prediction must also take into account the
ND and FD spectral differences that are present, even in the absence of
oscillations, due to pion decay kinematics and beamline geometry. This
is achieved using the {\em Beam Matrix} method. It utilizes the beam
simulation to derive a transfer matrix that relates the neutrinos in the
two detectors via their parent hadrons.  The ND reconstructed event
energy spectrum is translated into a flux by first correcting for the
simulated ND acceptance and then dividing by the calculated
cross-sections for each energy bin.  This flux is multiplied by the
transfer matrix to yield the predicted, unoscillated FD flux. After an
inverse correction for cross-section and FD acceptance, the predicted FD
visible energy spectrum is obtained.  The oscillation hypotheses are
then tested relative to this prediction. A distinct extrapolation
method, referred to as {\em ND Fit} was also applied to the data,
yielding similar results.

\begin{figure}[!htb]
  \begin{center}
  \includegraphics[height=4.5cm]{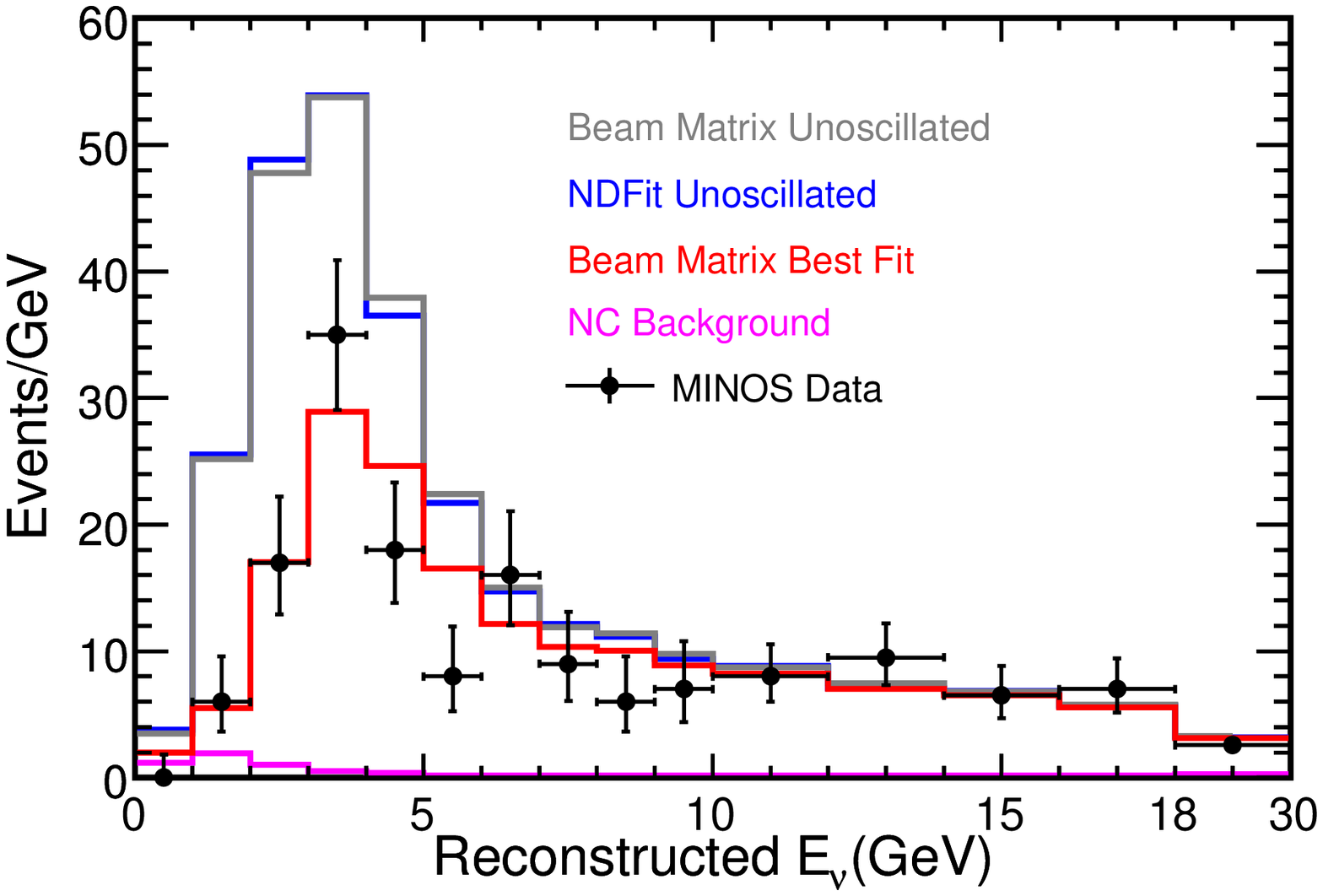}
  \includegraphics[height=4.5cm]{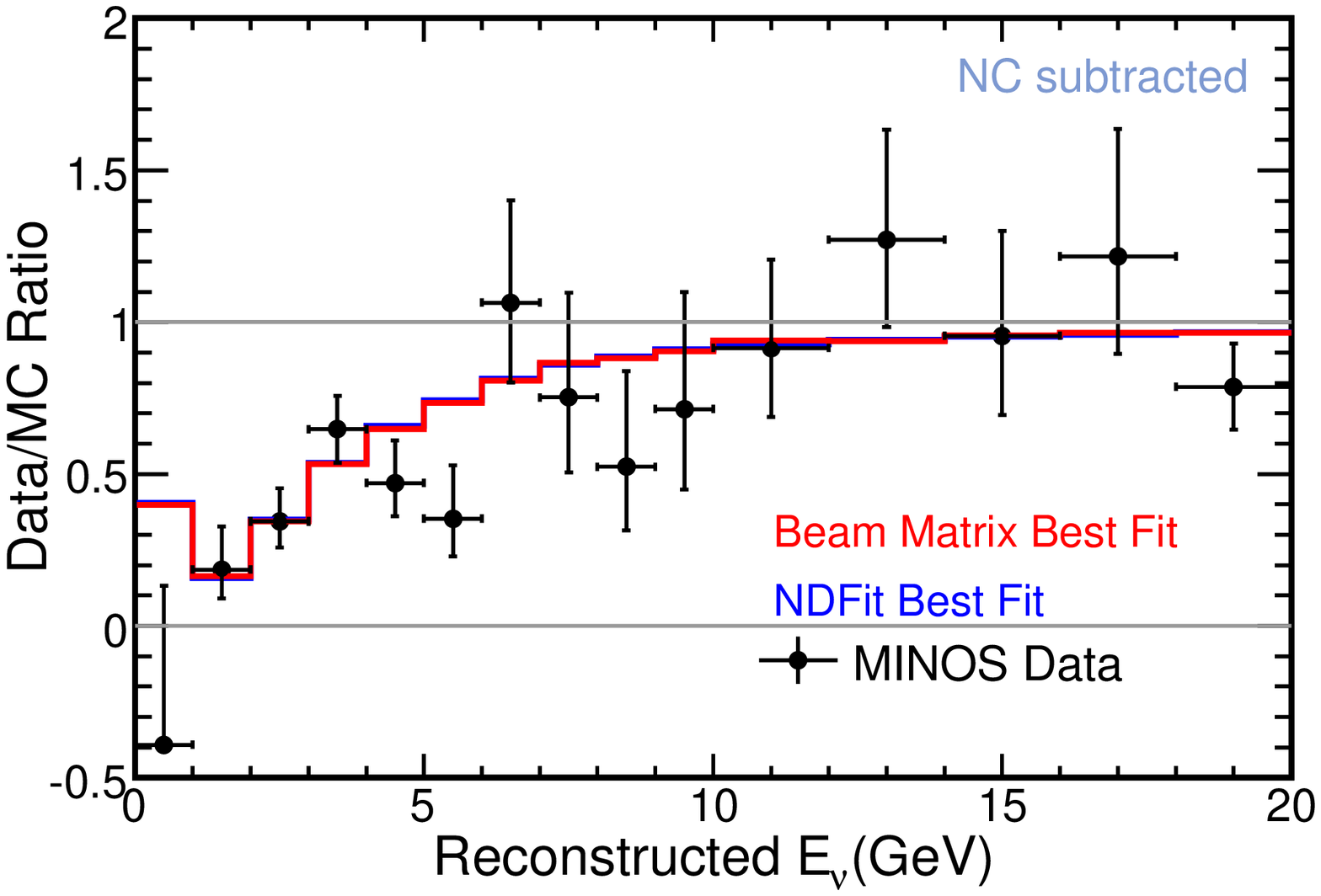}
  \caption{Comparison of the Far Detector spectrum with predictions for
    no oscillations for both analysis methods and for oscillations with
    the best fit parameters from the Beam Matrix extrapolation method
    (left canvas). The estimated NC background is also shown. The last
    energy bin contains events between 18-30\,GeV. The right canvas
    shows the ratio of data and best fit over the unoscillated predictions.
  }
  \label{fig:spectra}
\end{center}
\end{figure}

In total, 215 events are observed below 30\,GeV compared to
336.0$\pm$18.3(stat.)$\pm$14.4(syst.) events expected in the absence of
oscillations.  The systematic error is most relevantly due to NC
contamination, ND to FD normalization and the hadronic shower energy
scale.  In the region below 10\,GeV, 122 events are observed compared to
the expectation of 238.7$\pm$15.4$\pm$10.7.  The observed energy
spectrum is shown along with the predicted spectra for both
extrapolation methods in Figure\ \ref{fig:spectra}.

\subsection{Oscillation analysis}
\begin{figure}[!htb]
\begin{center}
\includegraphics[width=0.5\textwidth]{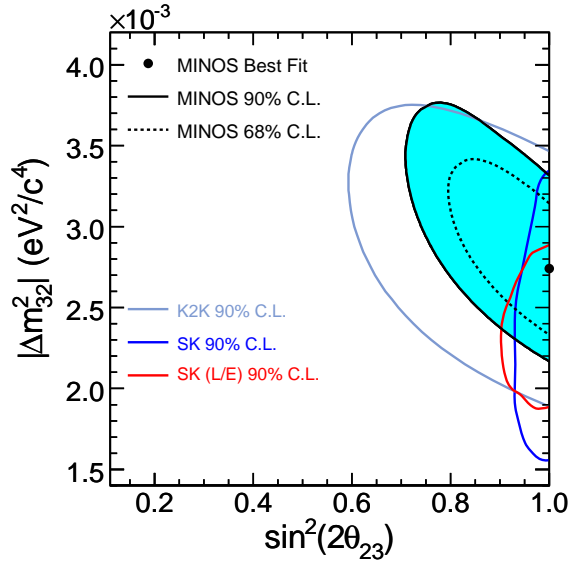}
\caption {Confidence intervals for the fit using the Beam Matrix method
  including systematic errors. Also shown are the contours from the
  previous highest precision experiments.
  \protect\cite{ref:osc1,ref:osc2,ref:osc5} }
\label{fig:contour} 
\end{center}
\end{figure}

Under the assumption that the observed deficit is due to \numutonutau{}
oscillations, a $\chi^2$ fit is performed to the parameters \dmsq{} and
\sintwo{} using the expression for the \numu{} survival probability:
\begin{equation}
  P(\nu_{\mu} \rightarrow \nu_{\mu})=1-\sin^2(2\theta_{23})
  \sin^2\!\left(\frac{\Delta m^2_{32} L}{4E}\right)
  \label{eq:osc}
\end{equation}
where $L$ is the distance from the target, $E$ is the neutrino energy,
and \dmsq{} is the atmospheric mass splitting.  The fit included the
systematic uncertainties mentioned above as nuisance parameters as well
as the small contribution from selected \nutau{} events produced in the
oscillation process. The resulting 68\% and 90\% confidence intervals
are shown in Figure\ \ref{fig:contour} as determined from
$\Delta$\cs=2.3 and 4.6, respectively.  The best fit parameter values
are:
\begin{equation}
|\Delta m^2_{32}|=(2.74\,^{+0.44}_{-0.26})\times 10^{-3}\,{\rm eV}^{2}/c^{4}
\end{equation}
and
\begin{equation}
\sin^2(2\theta_{23})>0.87
\end{equation} 
at 68\% C.L.\ with a fit probability of 8.9\%. At 90\% C.L.(2.31$<$\dmsq$<3.43)\times 10^{-3}$\,\evsq{}, and
\sintwo\,$>$\,0.78.  The data and best fit MC are shown in Figure\
\ref{fig:spectra}.

If the fit is not constrained to be within the physical region, the best
fit is at \dmsq{}=2.72\,$\times$\\$10^{-3}$\,\evsq{} and \sintwo{}=
1.01, with a decrease in \cs{} of 0.2.

It is expected that the systematic uncertainties will be reduced with
additional data. More details of this analysis are available in
\cite{ref:prl06}. An update on this result using 2.58$\times 10^{20}$
POT is expected during the Summer 2007.

\section{Neutrino time-of-flight analysis}
MINOS uses GPS synchronized clocks to timestamp neutrino interactions in
both detectors. This enables the measurement of the neutrino
time-of-flight over a distance of $734$\,km and thus the determination
of the neutrino velocity. Similar terrestrial experiments performed in
the past used much shorter baselines of $\sim500$\,m and higher beam
energies ($>30$\,GeV).

The time of each PMT hit is recorded by the detector's clock to a
precision of 18.8 ns in the Near Detector and 1.6 ns in the Far
Detector. The time of the earliest hit of each event is taken to be the
time of the neutrino interaction. The interaction times $t_1$ and $t_2$
in the two detectors are corrected for known offsets and delays, which
were determined using test stand measurements. In addition, the time of
the beam extraction signal is subtracted from both times. The beam
extraction signal has a fixed relation to the arrival of neutrinos in
the MINOS detectors. All times are therefore measured relative to this
reference.

\begin{figure}[tb]
\begin{center}
\includegraphics[width=0.6\textwidth]{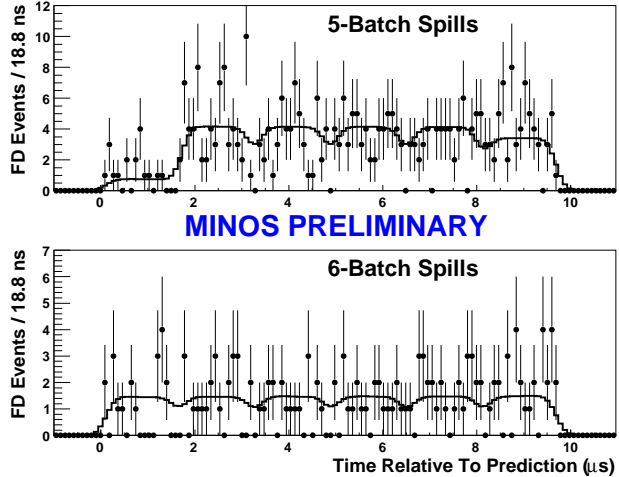}
\caption {Time distribution of FD events relative to prediction after
fitting the time-of-flight. The top plot shows events in 5-batch spills,
the bottom 6-batch spills. The solid lines show the normalized
prediction curves.}
\label{fig:TOF} 
\end{center}
\end{figure}

The NuMI beam pulse is not instantaneous, but has a duration of
$9.7\,\mu s$ with a five-batch or six-batch intensity profile depending
on the accelerator running mode. The Near Detectors measures this
time-intensity profile with neutrino interactions to high precision. The
measured time-intensity profile forms a probability density function
which is folded with a Gaussian distribution with a width of
$\sigma=150\,ns$ to account for the uncorrelated jitter of the two GPS
clocks. The resulting distribution $P(t)$ describes the predicted
arrival time distribution at the Far Detector (shown as a solid line in
Figure \ref{fig:TOF} for the five- and six-batch modes separately).

For the time-of-flight measurement, 473 neutrino-induced events in the
Far Detector were used. The time of each event was compared to the
predicted arrival time distribution. The time-of-flight $\tau$ was found
by maximizing an unbinned log-likelihood function:
\begin{equation}
L = \sum_i \ln P(t_2^i-\tau).
\end{equation}
The distribution of measured event times together with the predicted
distribution for the best fit $\tau$ is shown in Figure \ref{fig:TOF}.
The time-of-flight of neutrinos was measured to be 
\begin{equation}
  2449.223 \pm 0.032 \mathrm{(stat.)}\pm 0.064\mathrm{(syst.)}\,\mu
       \mathrm{s\qquad 68\%\ C.L.}
\end{equation}
Comparing to the MINOS baseline this translates to 
\begin{equation}
\frac{(v - c)}{c} = 5.1 \pm 2.9\,\mathrm{(stat.+sys.)} \times
10^{-5}\qquad 68\%\ \mathrm{C.L.}
\end{equation}
The systematic error is due to uncertainties on the timing delays and
offsets.

\section{Progress in Neutral Current analyses}
Analyses of neutral current neutrino interactions in MINOS are currently
in progress. Neutral current interactions are interesting for several
reasons. They form an important background to the \numutonutau
oscillation analysis described in these proceedings and their
cross-sections at the energies relevant to MINOS are not very well
known. Furthermore, an observed deficit of neutral current events at the
Far Detector could be evidence for light sterile neutrinos.

\begin{figure}[!htb]
\begin{center}
\includegraphics[width=0.49\textwidth]{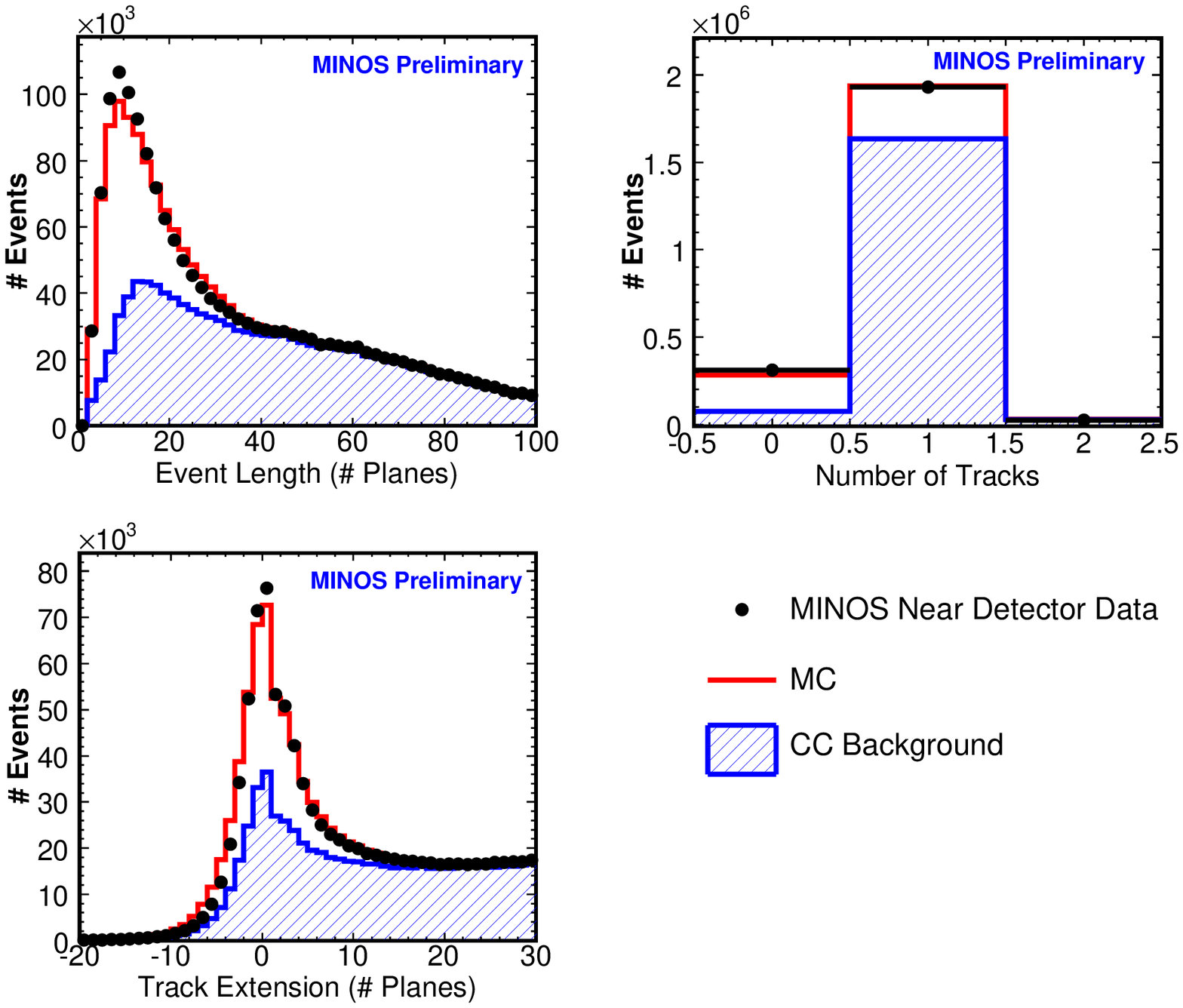}
\includegraphics[width=0.49\textwidth]{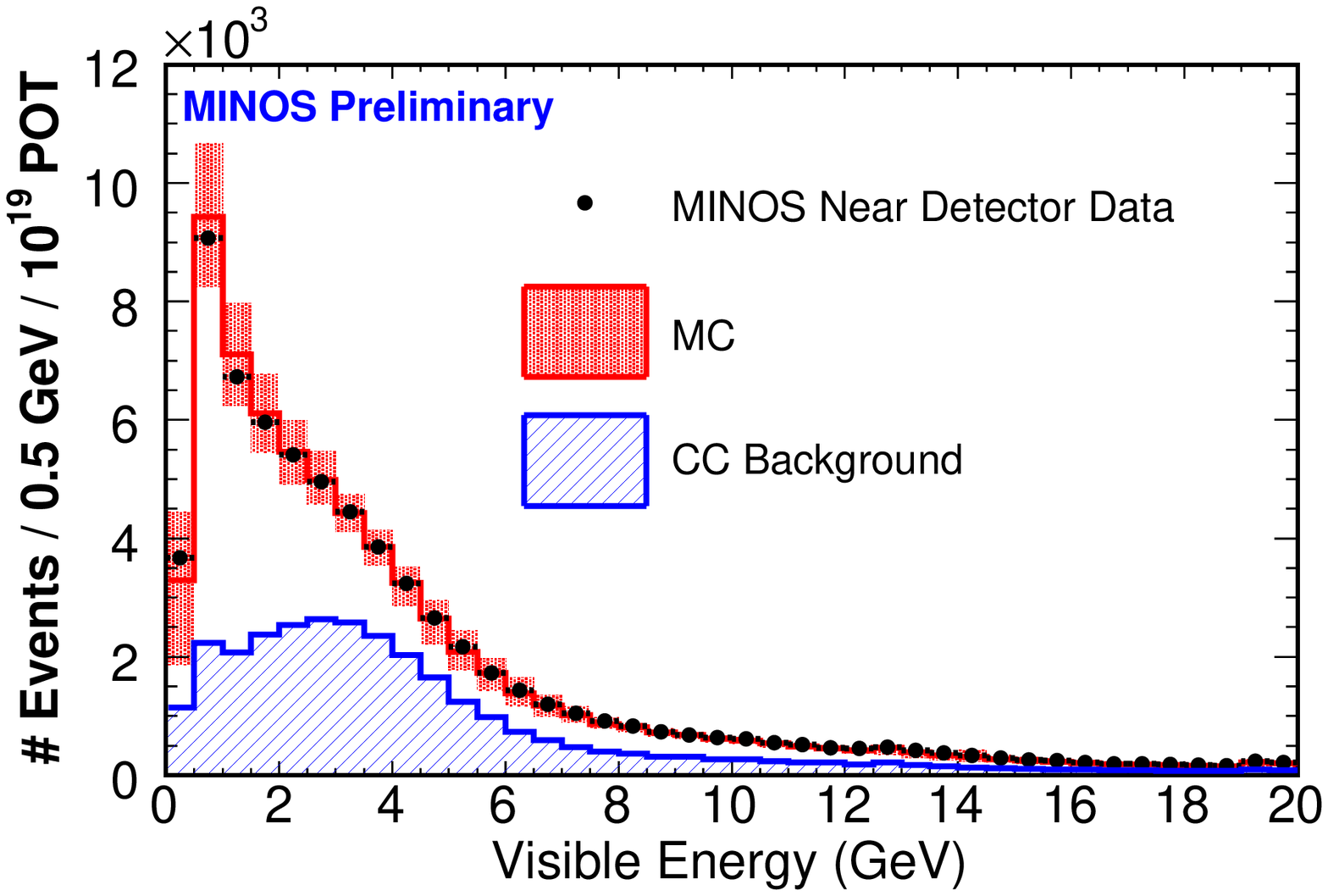}
\caption {Neutral current event selection variables in the Near Detector
(left). The data is shown as black markers, the Monte Carlo in red. The
blue distribution shows the charged current background. The right canvas
shows the reconstructed energy spectrum for selected neutral current
events. The Monte Carlo (red) is shown with systematic errors.}
\label{fig:ncSel} 
\end{center}
\end{figure}

Neutral current interactions are selected using three event quantities:
the event length, the number of tracks in the event and the {\em track
extension}. In events where both tracks and showers were found, the
track extension measures how much longer a reconstructed track is
compared to the hadronic shower it is accompanied by. Distributions of
the selection variables in Near Detector data and Monte Carlo are shown
in Figure \ref{fig:ncSel} (left canvas).

Using these variables, an energy spectrum of neutral-current-like events
was produced. This is shown in the right canvas of Figure\
\ref{fig:ncSel}. The Monte Carlo is shown as a red line with a
systematic error band; the background due to wrongly selected charged
current events is shown as a blue hatched distribution.  Within the
estimated systematic errors, due to flux, cross-section and energy scale
uncertainties, data and Monte Carlo agree well.

The extrapolation of these Near Detector results to the Far Detector and
a fit for oscillations to sterile neutrinos are currently being worked
on. Results from this analysis are expected later this year.

\section{Future Prospects}
In addition to the analyses reported here, MINOS is pursuing a
\numutonue oscillation analysis in order to measure or constrain the as
yet unknown mixing angle $\theta_{13}$. With its expected final
statistics, MINOS will potentially be able to improve on the current
best limit from the CHOOZ~\cite{ref:chooz} experiment.

Other areas of interest include appearance and disappearance
measurements of anti-neutrinos in the Far Detector as well as several
non-oscillation analyses using the large number of neutrino interactions
in the Near Detector.

\section*{Acknowledgments}
This work was supported by the US DOE; the UK PPARC; the US NSF; the
State and University of Minnesota; the University of Athens, Greece and
Brazil's FAPESP and CNPq. We are grateful to the Minnesota Department of
Natural Resources, the crew of the Soudan Underground Laboratory, and
the staff of Fermilab for their contributions to this effort.

\end{document}